\documentclass[conference]{IEEEtran}
\IEEEoverridecommandlockouts
\usepackage{cite}
\usepackage{amsmath,amssymb,amsfonts}
\usepackage{graphicx}
\usepackage{textcomp}

\usepackage{xcolor}
\def\BibTeX{{\rm B\kern-.05em{\sc i\kern-.025em b}\kern-.08em
    T\kern-.1667em\lower.7ex\hbox{E}\kern-.125emX}}

\usepackage{bm}

\newcounter{defcounter}
\usepackage{subfigure}

\usepackage{cases}
\usepackage{array}

\usepackage{empheq}

\usepackage{dsfont}

\newcommand{\mybold}{\text{\usefont{U}{bbold}{m}{n}1}}
\MakeRobust{\mybold}

\usepackage{bm}

\begin{document}
\title{Exploration of For-Purpose Decentralized Algorithmic Cyber Attacks in EV Charging Control}

%% To specify the authors when (number of affiliations <= 2)
\author{Mahan Fakouri Fard$^{\dagger}$, Xiang Huo$^{\dagger}$ and Mingxi Liu$^{\dagger}$% <-this % stops a space
% \thanks{This work was not supported by any organization}% <-this % stops a space
\thanks{$^{\dagger}$M. Fakouri Fard, X. Huo and M. Liu are with the Department of Electrical and Computer Engineering at the University of Utah, 50 S Central Campus Drive, Salt Lake City, UT, 84112, USA {\tt{\{mahan.fakourifard, xiang.huo, mingxi.liu\}@utah.edu}}.}}

%% To specify the authors when (number of affiliations > 2)
% \author{\IEEEauthorblockN{Author n.1\IEEEauthorrefmark{1},
% Author n.2\IEEEauthorrefmark{2},
% Author n.3\IEEEauthorrefmark{3}, 
% Author n.4\IEEEauthorrefmark{3} and
% Author n.5\IEEEauthorrefmark{4}}
% \IEEEauthorblockA{\IEEEauthorrefmark{1} Department Name of Organization A\\
% Name of the organization A,
% Address A\\ Emails if wanted}
% \IEEEauthorblockA{\IEEEauthorrefmark{2} Department Name of Organization B\\
% Name of the organization B,
% Address B\\ Emails if wanted}
% \IEEEauthorblockA{\IEEEauthorrefmark{3} Department Name of Organization C\\
% Name of the organization C,
% Address C\\ Emails if wanted}
% \IEEEauthorblockA{\IEEEauthorrefmark{4}Department Name of Organization D\\
% Name of the organization D,
% Address D\\ Emails if wanted}
% }

\maketitle

\begin{abstract}
 Distributed and decentralized multi-agent optimization (DMAO) algorithms enable the control of large-scale grid-edge resources, such as electric vehicles (EV), to provide power grid services. Despite its great scalability, DMAO is fundamentally prone to cyber attacks as it is highly dependent on frequent peer-to-peer communications. Existing cyber-security research in this regard mainly focuses on \emph{broad-spectrum} attacks aiming at jeopardizing the entire control system while losing the possibility of achieving specific attacking purposes. This paper, for the first time, explores novel \emph{for-purpose} algorithmic attacks that are launched by participating agents and interface with DMAO to achieve self-interest attack purposes. A decentralized EV charging control problem is formulated as an illustrative use case. Theoretical \emph{for-purpose} attack vectors with and without the stealthy feature are devised. Simulations on EV charging control show the practicability of the proposed algorithmic \emph{for-purpose} attacks and the impacts of such attacks on distribution networks.
\end{abstract}


\section{Introduction}
 The ever-growing electric vehicle (EV) penetration demands advanced control mechanisms to alleviate the negative impacts on distribution networks and increase power system flexibility \cite{optimizeEV_jin_2013}. Despite the related research progress, scalability and cyber security remain two major barriers to the large-scale deployment of EV charging control \cite{demanside_ipakchi_2011, towardefficient_cao_2017}. Control scalability ensures a timely engagement of significant power flexibility for grid service participation, while cyber security ensures data integrity as well as reliability of the distribution network. 

 Centralized control structures, due to the curse of scalability, are not suitable for large-scale EV charging control. In contrast, distributed and decentralized multi-agent optimization (DMAO) has presented outstanding scalability \cite{ADMMEV_Zhou_2021, DecentralizedEvSPDS_Liu} and is capable of integrating privacy-preserving measures \cite{admmPrivacy_zhang_2018}, thus are attracting growing attention. In \cite{ADMMEV_Zhou_2021, ADMMexchange_khaki_2019, ADMMplug_fan_2018}, the alternating direction method of multipliers (ADMM) was used to construct scalable distributed EV charging control schemes. In another research line, authors of \cite{DecentralizedEvSPDS_Liu,  huo2022two} developed the shrunken-primal-dual subgradient (SPDS) algorithms to construct decentralized EV charging control frameworks. Represented by ADMM and SPDS, DMAO algorithms have undoubtedly achieved  control scalability, however, must rely on iterative updates and frequent peer-to-peer communication, leading these algorithms prone to cyber attacks. 

%Despite the scalability advantages of DMAO algorithms, cyber security is a principal challenge in DMAO algorithm design. 
In the presence of malicious parties, once the transmitted information is obtained, altered, or jeopardized, cyber attackers can easily breach the entire operating system. Like other controllable grid-edge devices, EVs and their supply equipment are connected via the internet of things and are highly dependent on communication systems, leading EV charging control systems vulnerable to data manipulations \cite{loadfrquency_chen_2018}. Cyber security in DMAOs has recently attracted attention due to the vast use of decentralized and distributed control in industrial applications \cite{blockchain_sodhro_2020}. However, only a few works have attempted to investigate algorithmic cyber attacks that are integrated into DMAOs. The first attempt was made in \cite{CybersecADMM_Munsing_2018}, where the weaknesses of ADMM-based methods to various algorithmic attack vectors, including local problem distortion, noise injection, and coupling constraint distortion, were explored. However, only iterative noise injection attacks for convergence jeopardy were investigated through convexity-based  methods. Du \emph{et al.} in \cite{ADMMDos_Du_2019} investigated the impacts of data deception and  denial of service (DoS) on ADMM-based smart grid state estimation. Unfortunately, the proposed attacks have noticeable impacts on the system, thus lacking stealthiness. %Xu \emph{et al.} in \cite{admmOPF_xu_2021} deployed time delay attack and false data injection attack (FDIA) on ADMM-based optimal power flow problems. Further, he propose a monitoring system for the measurement variations based on the artificial neural network (ANN) algorithm. However, the presented result are preliminary, and efficiency of the proposed algorithm is not clear.
 
 Besides the limited advancement in algorithmic cyber attacks, two practical issues remain untouched in general cyber-security research. First, most existing work only focuses on \emph{broad-spectrum} attacks, e.g., DoS and noise injection, that aim at jeopardizing control stability \cite{towardefficient_cao_2017}, lowering the algorithm performance \cite{voltageregulation_liu_2022}, and/or preventing convergence \cite{optimalfdIA_liu_2017}. These attacks, unfortunately, cannot be adopted by internal attackers, i.e., algorithm participants, who want to achieve personal goals but still follow the algorithm. Second, most existing attack vectors have observable impacts on the system or the false data injected by attackers are noticeable \cite{ADMMDos_Du_2019}, making them easy to be detected by general detection methods  \cite{operationattack_chatterjee_2017,  hybridDetection_khan_2019}. Very few attempts were made to develop stealthy algorithmic attacks that are capable of concealing their impacts. In \cite{algorithmicattack_huang_2018}, an algorithmic attack was proposed to damage the critical power system infrastructure. A reachability-based
synthesis was developed to generate transient attacks that find attack parameters to avoid detection. Despite the detection avoidance performance, the proposed cyber-attack can only be used for overall system jeopardy rather than personal gain. %Besides, the proposed detection avoidance method only applies to specific topologies and data sets.  
 
 In this paper, we focus on the stealthy \emph{for-purpose} algorithmic attack that targets the DMAO iterations and can be imposed by algorithm participants. The contribution of this paper is three-fold: (1) A novel \emph{for-purpose} algorithmic cyber attack, which allows attackers to  manipulate the DMAO algorithm to gain sophisticated personal benefits, is explored; (2) Two practical self-interest attack vectors in EV charging control are investigated. Their corresponding impacts on the distribution network are analyzed; (3) A novel mechanism is developed to grant stealthy features to the proposed algorithmic attacks. The proposed methods are rather general for DMAO algorithms -- EV charging control is used in this paper for better illustration.

\section{EV Charging Control Scheme}
\subsection{Distribution network and EV charging model}
This paper adopts the LinDisFlow model \cite{baran1989optimal} to represent a linear relationship between EV charging power and squared nodal voltage magnitudes. In a distribution network with $n$ buses, at time $t$, the LinDistFlow model gives
\begin{equation}
\bm{V}(t)=\bm{V}_{0}-2 \bm{R} \bm{P}(t)-2 \bm{X} \bm{Q}(t),
\label{eq:1}
\end{equation}
where $\bm{P}(t) \in \mathbb{R}^{n}$ and $\bm{Q}(t) \in \mathbb{R}^{n}$ denote the real and reactive power consumption of all buses, respectively, $\bm{V}(t) \in \mathbb{R}^{n}$ contains the squared voltage magnitudes of all buses, and $\bm{V}_0 = V_0^2 \bm{1}_n \in \mathbb{R}^{n}$ denotes the slack constant voltage magnitude vector with $V_0$ being the voltage magnitude at the feeder head. $\bm{R}$ and $\bm{X}$ are the adjacency matrices defined as 
\begin{equation} \label{3}
\begin{aligned}
\bm{R} &\in \mathbb{R}^{n \times n}, \quad R_{\nu \kappa}=\sum_{(\nu, \kappa) \in \mathbb{E}_{\nu} \cap \mathbb{E}_{\kappa}} 
r_{\nu\kappa}, \\
\bm{X}  &\in \mathbb{R}^{n \times n}, \quad 
X_{\nu\kappa}=\sum_{(\nu,\kappa) \in \mathbb{E}_{\nu} \cap \mathbb{E}_{\kappa}} 
x_{\nu\kappa},
\end{aligned}
\end{equation}
where $r_{\nu,\kappa}$ and $x_{\nu,\kappa}$ are the resistance and reactance of  line $(\nu,\kappa)$, respectively, and $\mathbb{E}_{\nu}$ and $\mathbb{E}_{\kappa}$ are the line sets connecting the feeder head to bus $\nu$ and $\kappa$, respectively \cite{farivar2013equilibrium}. %, i.e., the elements $R_{ij} (X_{ij})$ in the adjacency matrices are obtained through the resistance (reactance) of the common path of  bus $i$ and bus $j$ leading back to the feeder head \cite{zhou2020gradient}.

The power consumption at each node consists of the baseline and EV charging load. Assuming the EVs only consume real power, at node $l$, we have $p_l(t)=p_{l,b}(t)+p_{l,EV}(t)$ and $q_l(t)=q_{l,b}(t)$, where $p_{l,b}(t)$, $q_{l,b}(t)$, and $p_{l,EV}(t)$ denote the real baseline power, reactive baseline power, and EV charging power, respectively. Let $\bm{V}_b(t)$ denote the squared voltage drop caused by the baseline load, \eqref{eq:1} can be rewritten as
%\begin{subequations} \begin{align}
 %   &p_l(t)=p_{l,b}(t)+p_{l,EV}(t)\\
  %  &q_l(t)=q_{l,b}(t)
   % \end{align}
%\end{subequations}  
\begin{equation}
    \bm{V}(t)=\bm{V}_0 - \bm{V}_b(t) - 2 \bm{R}p_{EV}(t),
\end{equation} where $p_{EV}(t)=[p_{1,EV}(t) ~ p_{2,EV}(t)~ \cdots~  p_{n,EV}(t)]^{\mathsf{T}}$. Suppose $s_l$ EVs are connected to node $l$, we have $p_{l,EV}(t)= \sum_{\hat{l}=1}^{s_{l}}\tilde{P}_{l,\hat{l}}c_{l,\hat{l}}(t),$
%\begin{equation}
 %   p_{l,EV}(t)= %\sum_{\hat{l}=1}^{s_{l}}\tilde{P}_{l,\hat{l}}c_{l,\hat{l}}(t), 
%\end{equation}
where $c_{l,\hat{l}}(t)$ and $\tilde{P}_{l,\hat{l}}$ denote the charging rate and maximum charging power of the $\hat{l}$th EV connected to node $l$. Re-indexing $c_{l,\hat{l}}$ and $\tilde{P}_{l,\hat{l}}$ with $i=1,\dots,s$ by following the ascending orders of $l$ and $\hat{l}$, and defining $\bm{G}=\oplus_{l=1}^n \bm{G}_l \in \mathbb{R}^{n\times s}$ and $\tilde{\bm{P}}=\oplus_{i=1}^s \tilde{P}_i \in \mathbb{R}^{s\times s}$, we have
\begin{equation}
    \bm{V}(t)=\bm{V}_0 - \bm{V}_b(t) - 2 \bm{R}\bm{G}\tilde{\bm{P}}\bm{C}(t),
\end{equation} where $\bm{G}_l=\bm{1}_{s_l}^{\mathsf{T}}$ is the charging power aggregation vector, $\bm{C}(t)=[c_1(t) ~ c_2(t) ~ \cdots ~ c_s(t)]^{\mathsf{T}} \in \mathbb{R}^s$, and $\oplus$ denotes matrix direct sum. Further let $\bm{D}\in\mathbb{R}^{n\times s}$ denote $- 2 \bm{R}\bm{G}\tilde{\bm{P}}$, $\bm{y}_d(t)$ denote $\bm{V}_0 - \bm{V}_b(t)$ and $\bm{y}(t)$ denote $\bm{V}(t)$, we have
\begin{equation}
    \bm{y}(t)= \bm{y}_d(t) + \bm{D}\bm{C}(t).
\end{equation}

Let $SOC_{i,ini}$ and $SOC_{i,des}$ denote the initial and the desired SOC of the $i$th EV, respectively, and $\hat{E}_i$ denotes the battery capacity of $i$th EV. Then the total battery energy required by the $i$th EV is $E_{i,req}=\hat{E}_i(SOC_{i,des}-SOC_{i,ini}).$

%Let define the system state $u_i(t)$ as the energy remaining to be charged in $i$th EV in order to achieve the total required energy $ E_{i,req}$. Now the charging dynamic can be written as
%\begin{equation}
%    u_i(t+1)=u_i(t)+ B_i c_i(t)
%\end{equation} where $B_i=-\eta_i\Delta t\tilde{P_i}$ and $\eta_i$ is the charging efficiency. Suppose $i$th EV plug in to the charger at time $t_i$, and the $T_i$ is the maximum wait time for reaching the desired SOC. Then we should have \begin{equation}
%    u_i(t_i+T_i)=0.
%\end{equation} Let $s$ denotes the number of EVs and $\bm{B}_{i,c}\in \mathbb{R}^s$ denotes the the $i$th column of the matrix $\bm{B}=\oplus_{i=1}^s \bm{B}_i$ then, we have
%\begin{equation}
%    u(t+1)=u(t) + \sum_{i=1}^s \bm{B}_{i,c} c_i(t),
%\end{equation} where $\bm{u}(t)=[u_1(t) ~~ u_2(t) ~~ \dots ~~ u_s(t)]^\intercal \in \mathbb{R}^s$ and . Therefore, at time $t$, we can write the system dynamics as
%\begin{subequations}
%    \begin{align}
%        &\bm{u}(t+1)=\bm{u}(t)+\bm{B}\bm{C}(t) \label{12a}\\
%        &y(t)=y_d(t)+\bm{D}\bm{C}(t). \label{12b}
%    \end{align}
%\end{subequations}.

\subsection{Valley filling problem}

The goal of valley filling is to use the aggregated EV charging power to  fill the overnight electricity use valley. The valley-filling problem can be modeled as an optimal power flow problem that minimizes the variance of the aggregated total load. Let $T$ be the valley filling period, then the charging profile of the $i$th EV is represented as $ \bm{\mathcal{C}}_i=[c_i(t)~c_i(t+1)~\cdots~c_i(t+T-1)]^{\mathsf{T}} \in \mathbb{R}^T.$ 
%\begin{equation}
%    \bm{\mathcal{C}}_i=[c_i(t)~c_i(t+1)~\cdots~c_i(t+T-%1)]^{\mathsf{T}} \in \mathbb{R}^T.
%\end{equation}
In a centralized fashion, let $\bm{\mathcal{C}} = [\bm{\mathcal{C}}_1^{\mathsf{T}}  \cdots \bm{\mathcal{C}}_s^{\mathsf{T}} ]^{\mathsf{T}} \in \mathbb{R}^{sT}$ denote the collection of all EVs' charging profiles, the valley-filling problem is formulated as
\begin{subequations} \label{main problem}
    \begin{align}
    &\min_{\bm{\mathcal{C}}}~\mathcal{F}(\bm{\mathcal{C}})=\frac{1}{2}\left\|\bm{P}_b+\sum_{i=1}^{s}\tilde{P_i}\bm{\mathcal{C}}_i\right\|_2^2 \label{3a}\\
    & ~~ \text{s.t.} ~~\bm{\mathcal{C}}_i\in\mathbb{C}_i, ~ \forall i \in 1,2,...,s, \label{3b}\\
    &~~~~~~~\bm{\mathcal{Y}}_b-\sum_{i=1}^n\bm{\mathcal{D}}_i\bm{\mathcal{C}}_i\leq \bm{0},
    \label{3c}
    \end{align}
\end{subequations}
where $\bm{P}_b = [P_b(t)~P_b(t+1)~~\cdots ~~ P_b(t+T-1)]^{\mathsf{T}}\in\mathbb{R}^T$ is the aggregated baseline load profile of the entire distribution network.  The constraint set $\mathbb{C}_i$ guarantees the $i$th EV can be charged to the desired SOC by the end of the valley filling period, which takes the form of
\begin{equation} \label{primal constraints}
   \mathbb{C}_i:=\{\bm{\mathcal{C}}_i| \bm{0} \leq \bm{\mathcal{C}}_i \leq \bm{1}, E_{i,req}-\bm{\hat{B}}_{i,l}\bm{\mathcal{C}}_i=0\},
    \end{equation}
    where $\bm{\hat{B}}_{i,l}= \bm{1}_s\bm{B}_{i,l}$, $\bm{B}_{i,l}=[\bm{B}_{i,c}~\bm{B}_{i,c}~\cdots~\bm{B}_{i,c}]\in \mathbb{R}^{s\times T}$, $\bm{B}_{i,c}\in \mathbb{R}^s$ denotes the the $i$th column of the matrix $\bm{B}=\oplus_{i=1}^s B_i$, $B_i=-\eta_i\Delta t\tilde{P_i}$, $\eta_i$ is the charging efficiency, $\Delta t$ is the sampling time, and $\bm{1}_s=[1\cdots 1]\in \mathbb{R}^{1\times s}$. Eqn. \eqref{3c} ensures all nodal voltage magnitudes stay above the lower bound, where  $\bm{\mathcal{Y}}_b$ denotes $\underline{v}^2\bm{V_0}-\bm{\mathcal{Y}}_{d}$, $\bm{\mathcal{Y}}_d=[y_d(t)~y_d(t+1)~\cdots~y_d(t+T-1)]^{\mathsf{T}}\in \mathbb{R}^{nT}$, $\underline{v}$ is the bus voltage magnitude lower bound, $\bm{\mathcal{D}_i}=D_i\oplus D_i\dots\oplus D_i\in\mathbb{R}^{sT\times T}$ denotes the mapping between EV charging power and the nodal voltage magnitudes, and $\bm{D}=[D_1~D_2~\cdots~D_s]$.  

\subsection{Decentralized EV charging control}
To achieve control scalability, we adopt SPDS \cite{DecentralizedEvSPDS_Liu} to solve \eqref{main problem} in a decentralized way. With the relaxed Lagrangian
of problem \eqref{main problem} defined as 
\begin{equation}
    \mathcal{L}(\bm{\mathcal{C}},\bm{\lambda}) = \mathcal{F}(\bm{\mathcal{C}}) + \bm{\lambda}^{\mathsf{T}} \big(\bm{\mathcal{Y}}_b-\sum_{i=1}^n\bm{\mathcal{D}}_i\bm{\mathcal{C}}_i\big), 
\end{equation}
 each EV iteratively updates the primal variable by following 
\begin{equation} \label{primal update}
     \bm{\mathcal{C}}_i^{(k+1)} {=} \Pi_{\mathbb{C}_i}\left(\frac{1}{\tau_\mathcal{C}}\Pi_{\mathbb{C}_i}\left(\tau_\mathcal{C}\bm{\mathcal{C}}_i^{(k)}{-}\alpha_{i,k}\nabla_{\bm{\mathcal{C}}_i}\mathcal{L}(\bm{\mathcal{C}}^{(k)},\bm{\lambda}^{(k)})\right)\right),
    \end{equation}
    where $\tau_\mathcal{C}$ is the primal shrinking parameter and $\alpha_{i,k}$ is the primal update step size. Similarly, SPDS iteratively updates the dual variable by following
    \begin{equation} \label{dual update}
     \bm{\lambda}^{(k+1)} = \Pi_{\mathbb{D}}\bigg(\frac{1}{\tau_\lambda}\Pi_{\mathbb{D}}\big(\tau_\lambda \bm{\lambda}^{(k)}+\beta_{k}\triangledown_{\lambda}\mathcal{L}(\bm{\mathcal{C}}^{(k)},\bm{\lambda}^{(k)})\big)\bigg),
\end{equation}
where $\bm{\lambda}\in \mathbb{R}^{nT}$ is the dual variable associated with \eqref{3c}, $\tau_\lambda$ is the dual shrinking parameter, and $\beta_{i,k}$ is the dual update step size. 
%\begin{equation} \label{dual constraints}
%    \mathbb{D}:=\{\bm{\lambda}|\bm{\lambda} \ge 0, \|\bm{\lambda}\|_2\leq d_\lambda \}.
%\end{equation}
Under Slater condition, $\mathbb{D}$ is non-empty \cite{ koshal2011siam}. %Further, based on \cite{koshal2011siam}, at the Slater point of \eqref{main problem}, the strong duality holds, there exists a primal solution, and $d_\lambda$ can be obtained a \emph{priori}. 

By implementing SPDS in EV charging control, individual EV chargers only need to share their own $\bm{\mathcal{C}}_i^{(k)}$ with the system operator. The system operator computes the Lagrange gradient and $\bm{\lambda}^{(k)}$ and broadcasts them to all EVs. This process will continue until the tolerance $\|\bm{\mathcal{C}}^{(k+1)}-\bm{\mathcal{C}}^{(k)}\|_2$ drops below a threshold. The convergence of SPDS is proved in\cite{DecentralizedEvSPDS_Liu}. 

%It can be concluded from \cite{DecentralizedEvSPDS_Liu} that The shrinkage-expansion and the two-tier projection in the primal and the dual updates guarantee the convergency without any regularization. 

\section{For-Purpose Algorithmic Cyber Attacks}

As aforementioned, one common issue of existing cyber-attack studies is that the goals of the attackers are not self-beneficial or rather realistic, and because of the noticeable impacts, they are more likely detectable. To resolve this issue, we will devise a new algorithmic attack -- \emph{for-purpose} cyber attack, where the attacker injects deliberate data into the DMAO iterations to gain self-beneficial results without affecting algorithm convergence or making noticeable changes from the truly optimal operation. The concept of the presented cyber attack enables multiple attack scenarios.  Specifically, in EV charging control, EVs can inject sophisticated data into their own and/or others' communication channels to achieve personal benefits while not affecting algorithm convergence. However, due to the page limit, we only consider the scenario where the attackers only manipulate their own data.

 %With some knowledge of the operating algorithm, a persistent \emph{for-purpose attack} can change the optimal solution of the main problem, which is more desired for the attacker. It should be noted that the attackers want to remain undetected; hence, they would not try to impact the monitored nodal voltages or bus loads significantly. In this paper we focus on attacks that do not jeopardize the convergence of the algorithm but participate in the problem and benefits from it.
 
%In the next section, we will introduce an algorithmic attack model that can be applied to the SPDS algorithm, in which, first, the attackers can gain personal benefits and, second, they can stay out of detection at the same time.        

\subsection{Self-interest algorithmic attack vectors} \label{sec_self_attack}

Suppose the $i$th EV wants to pursue a self-interest objective represented as $\omega_1\mathcal{G}(\bm{\mathcal{C}}_i)$, where $\omega_1>0$ denotes the power of the self-interest attack. We have the following theorems.

\noindent \textbf{Theorem 1:} An internal attacker who follows SPDS algorithm can deviate the optimal solution of the problem in \eqref{main problem} towards its convex interest function $\mathcal{G}(\bm{\mathcal{C}}_i)$ by only modifying its local primal update direction by $\omega_1\nabla_{\bm{\mathcal{C}}_i}\mathcal{G}(\bm{\mathcal{C}}_i)$, i.e.,
\begin{equation} \label{attack primal update}
     \bm{\mathcal{C}}_i^{(k+1)} = \Pi_{\mathbb{C}_i}\bigg(\frac{1}{\tau_\mathcal{C}}\Pi_{\mathbb{C}_i}\big(\tau_\mathcal{C}\bm{\mathcal{C}}_i^{(k)}-\alpha_{i,k}\nabla_{\bm{\mathcal{C}}_i}\tilde{\mathcal{L}}(\bm{\mathcal{C}}^{(k)},\bm{\lambda}^{(k)})\big)\bigg),
    \end{equation} 
 where  $\tilde{\mathcal{L}}(\bm{\mathcal{C}}^{(k)},\bm{\lambda}^{(k)})= \mathcal{L}(\bm{\mathcal{C}}^{(k)},\bm{\lambda}^{(k)})+\omega_1\mathcal{G}(\bm{\mathcal{C}}_i)$ while the algorithm convergence is guaranteed. \hfill $\blacksquare$ 

\noindent\textbf{Proof:} Locally modifying the $i$th EV's primal update as in \eqref{attack primal update} is equivalent to modifying the problem in \eqref{main problem} to
\begin{subequations} \label{attack problem}
    \begin{align}
    &\min_{\bm{\mathcal{C}}}~\mathcal{F}(\bm{\mathcal{C}}) +\omega_1\mathcal{G}(\bm{\mathcal{C}}_i)  \label{8a}\\
    & ~~ \text{s.t.} ~~\eqref{3b},~\eqref{3c}.
    \end{align}
\end{subequations}
Therefore, following the convergence proof of SPDS \cite{DecentralizedEvSPDS_Liu}, as long as $\mathcal{G}(\bm{\mathcal{C}}_i)$ is convex and $\omega_1>0$, the algorithm convergence is guaranteed. Because the voltage constraints remain unchanged, the converged results satisfy the global voltage requirements. Since, in each iteration, the $i$th EV's primal update is re-directed to honor the descending direction of $\mathcal{G}(\bm{\mathcal{C}}_i)$, the converged results are in favor of the $i$th EV. \hfill $\square$

\noindent\textbf{Theorem 2:} Given $\mathcal{G}(\bm{\mathcal{C}}_i) \geq 0$ for all feasible $\bm{\mathcal{C}}_i$, the optimal solution for the problem in \eqref{attack problem} differs from the optimal solution for problem \eqref{main problem} and the difference is bounded by $\omega_1  \max\{\mathcal{G}(\bm{\mathcal{C}}), \forall \bm{\mathcal{C}_i} \in  \mathbb{S}\}$. \hfill $\blacksquare$

\noindent \textbf{Proof:} Let $\bm{\mathcal{C}}^* \in \mathbb{R}^{sT\times 1}$ denote the optimal solution of the attack-free problem in \eqref{main problem} with feasible region $\mathbb{S}$, then 
\begin{equation} \label{one}
    \mathcal{F}(\bm{\mathcal{C}}^*)\leq \mathcal{F}(\bm{\mathcal{C}}), ~\forall \bm{\mathcal{C}}\in \mathbb{S}.
\end{equation}
Further, let $\hat{\bm{\mathcal{C}}}$ be the optimal solution of the attacked problem \eqref{attack problem} with the same feasible region $\mathbb{S}$. Since for all feasible $\bm{\mathcal{C}}_i$, $\mathcal{G}(\bm{\mathcal{C}}_i) \geq 0$ and $\omega_1>0$, it follows that
\begin{equation} \label{four}
    \mathcal{F}(\hat{\bm{\mathcal{C}}})\leq \mathcal{F}(\hat{\bm{\mathcal{C}}})+\omega_1 \mathcal{G}(\hat{\bm{\mathcal{C}}}_i).
\end{equation}
Therefore, it can be readily derived that
\begin{equation} \label{six}
    \mathcal{F}(\bm{\mathcal{C}}^*) \leq \mathcal{F}(\hat{\bm{\mathcal{C}}}) \leq \mathcal{F}(\hat{\bm{\mathcal{C}}}) + \omega_1\mathcal{G}(\hat{\bm{\mathcal{C}}}_i),
\end{equation}
indicating that the optimal value of the attack-free problem \eqref{main problem} is always no greater than the optimal value of the attacked problem \eqref{attack problem}. According to multi-objective optimization theories, a Pareto-optimal set is the set of all optimal solutions such that no other solution can improve one objective function without deteriorating another \cite{multiobj_ngatchou_2005}. Therefore, based on \eqref{six}, the Pareto-optimal set for $\mathcal{F}(\bm{\mathcal{C}}) + \omega_1\mathcal{G}(\bm{\mathcal{C}}_i)$ will always contain the Pareto-optimal set for $\mathcal{F}(\bm{\mathcal{C}})$. Hence, $\bm{\mathcal{C}}^*$ and $\bm{\mathcal{\hat{C}}}$ cannot belong to the same Pareto-optimal set, and they must be different. Note that the upper bound for the total deviation from the original optimal solution after the attack is dependent on the attack function and power, $\mathcal{{G}}(\bm{\mathcal{C}}_i)$ and $\omega_1$. Generally, based on \eqref{six}, this upper bound can be calculated as
\begin{equation}
  \mathcal{F}(\bm{\mathcal{C}}^*)-\mathcal{F}(\hat{\bm{\mathcal{C}}}) \leq \omega_1\mathcal{G}(\hat{\bm{\mathcal{C}}}_i)\leq  \omega_1  \max\{\mathcal{G}(\bm{\mathcal{C}}_i), \forall \bm{\mathcal{C}} \in  \mathbb{S}\}.
\end{equation}
\hfill $\square$

The assumption of non-negative self-interest objective in \textbf{Theorem 2} is generally true for most attacking purposes. In what follows, we present two possible scenarios. Linear $\mathcal{G}(\bm{\mathcal{C}}_i)$ that may violate this assumption will be studied in future work.

%The attack vector $\mathcal{G}_i(\mathcal{C}_i)$ can take various forms to achieve different personal benefits. 

\subsubsection{Smooth-charging attack}   
In this case, the $i$th EV aims to converge to a relatively smoother charging profile to preserve the battery's state of health (SOH). To this end, the $i$th EV can modify its local primal update to
\begin{equation}
    \begin{aligned}
    \bm{\mathcal{C}}_i^{(k+1)} = & \Pi_{\mathbb{C}_i}\bigg(\frac{1}{\tau_\mathcal{C}}\Pi_{\mathbb{C}_i}\big(\tau_\mathcal{C}\bm{\mathcal{C}}_i^{(k)}-\alpha_{i,k}\nabla_{\bm{\mathcal{C}}_i}{\mathcal{L}}(\bm{\mathcal{C}}^{(k)},\bm{\lambda}^{(k)}) \\
    & ~~~~~~~~~~~~~~~~~-2\alpha_{i,k}\omega_1\bm{\mathcal{C}}_i^{(k)} \big)\bigg).
    \end{aligned}
\end{equation}
%\begin{equation} \label{smooth attack primal update}
%     \bm{\mathcal{C}}_i^{(k+1)} = \Pi_{\mathbb{C}_i}\bigg(\frac{1}{\tau_\mathcal{C}}\Pi_{\mathbb{C}_i}\big(\tau_\mathcal{C}\bm{\mathcal{C}}_i^{(k)}-\alpha_{i,k}\triangledown_{\bm{\mathcal{C}}_i}\tilde{\mathcal{L}}(\bm{\mathcal{C}}^{(k)},\bm{\lambda}^{(k)})\big)\bigg)
%    \end{equation} 
%    \begin{equation} \label{smooth attack lagrang}
%       \tilde{\mathcal{L}}(\bm{\mathcal{C}}^{(k)},\bm{\lambda}^{(k)})= \mathcal{L}(\bm{\mathcal{C}}^{(k)},\bm{\lambda}^{(k)})+\omega_1\|\bm{\mathcal{C}}_i\|_2^2
%    \end{equation}
As shown in the proof of \textbf{Theorem 1}, if the $i$th EV injects $2\omega_1\bm{\mathcal{C}}_i^{(k)}$ in every iteration, the overall EV charging control problem is equivalent to adding $\omega_1\|\bm{\mathcal{C}}_i\|_2^2$ to the objective function to flatten the $i$th EV's charging profile. The impacts of the charging profile change of the $i$th EV on the valley-filling performance will be compensated for by other EVs.

\subsubsection{Rush-charging attack}
In this scenario, the $i$th EV aims to charge as soon as possible. To this end, it needs to inject $2\omega_1 \bm{A}^{\mathsf{T}} \bm{A}\bm{\mathcal{C}}_i^{(k)}$ at each primal update iteration, where $\bm{A}\in \mathbb{R}^{T\times T}$ is a diagonal matrix with each element
\begin{equation}
A_{\hat{t},\hat{t}}=\left\{ \begin{array}{ll}
m,&~ \text{if } \hat{t} \leq t_d \\
M,&~ \text{otherwsie}
\end{array}\right.
\end{equation}
Herein, $0<m \ll M$ and $t_d$ denotes the attacker's desired termination time to reach $SOC_{i,des}$. According to \textbf{Theorem 1}, this is equivalent to adding $\omega_1\|\bm{A}\bm{\mathcal{C}}_i\|_2^2$ to the objective function of problem \eqref{main problem}.
Entries in $\bm{A}$ with smaller values will force the corresponding elements in $\bm{\mathcal{C}}_i$ to be maximized, and \emph{vice versa}, to achieve the rush-charging goal. 
 
\subsection{Stealthy for-purpose algorithmic attack vector}

Though the attack vectors developed in Section \ref{sec_self_attack} can assist individual EVs to attain self-interest objectives, according to \textbf{Theorem 2}, the deviations in the objective value, reflected by the impaired valley filling performance, may inform the system operator about the existence of cyber attacks. Therefore, in order to remain stealthy, it is critical to develop an attacking mechanism to minimize the deviation of the post-attack results from the true optimal solution. To this end, it is ideal for the $i$th EV to launch an attack that equivalently converts the EV charging control problem \eqref{main problem} to 
\begin{subequations} \label{stealth attack problem_dumy}
    \begin{align}
    &\min_{\bm{\mathcal{C}}}~\mathcal{F}(\bm{\mathcal{C}}) +\omega_1\mathcal{G}(\bm{\mathcal{C}}_i) + \omega_2\|\bm{\mathcal{C}}-{\bm{\mathcal{C}}}^*\|_2^2  \label{8a}\\
    & ~~ \text{s.t.} ~\eqref{3b},~\eqref{3c},
    \end{align}
\end{subequations}
where $\bm{\mathcal{C}}^*$ denotes the true optimal solution of problem \eqref{main problem}. %Let $\bm{I}_i \in \mathbb{R}^{sT\times T}$ denote a block matrix whose $i$th block is an identity matrix $\bm{I} \in \mathbb{R}^{T\times T} $ and others are zeros, therefore we have 
%$\bm{\mathcal{C}} = \sum_{i=1}^{s}   \bm{\mathcal{C}}_i \bm{I}_i$. 

To realize this, the $i$th EV needs to inject $\omega_1\nabla_{\bm{\mathcal{C}}_i}\mathcal{G}(\bm{\mathcal{C}}_i^{(k)})+2\omega_2\bm{I}_i^{\mathsf{T}}(\bm{\mathcal{C}}^{(k)}-\bm{\mathcal{C}}^{(\ell)})$ into its primal update, where  $\bm{I}_i \in \mathbb{R}^{sT\times T}$ denotes a block matrix whose $i$th block is an identity matrix $\bm{I} \in \mathbb{R}^{T\times T} $ and others are zeros. Two challenges exist: First, the attacker is not able to obtain $\bm{\mathcal{C}}^*$ after launching the attacks. Second, the second term in the injected malicious data requires knowledge of other EVs' intermediate decision variables. Herein, for the purpose of exploring the existence of stealthy algorithmic attacks, we assume that an attacker is capable of wiretapping the communication channels between other EVs and the system operator, which resolves the second challenge. This assumption will be lifted in future work. %However, The attacker can have find a close estimation of the optimal solution of the main problem and try to deviate the after attack optimal solution as close as possible to it. 

Though it is impossible to obtain $\bm{\mathcal{C}}^*$ of the attack-free problem, it is possible for the $i$th EV to make an estimation. In the convergence of DMAO algorithms, with an overwhelming possibility, all decision variables converge or meet the stopping criterion at the same time. Based on this, the $i$th EV (the attacker) first pre-determines a threshold $\epsilon_{att}^*$ for the difference between two consecutive iterations $\epsilon_{att}^{(\ell)}=\|\bm{\mathcal{C}}_i^{(\ell+1)}-\bm{\mathcal{C}}_i^{(\ell)}\|_2$. At the beginning of valley filling, the $i$th EV allows the algorithm to run normally without launching any attack. As the iteration goes, once $\epsilon_{att}^{(\ell)}$ drops below $\epsilon_{att}^*$ in the $\ell$th iteration, the $i$th EV regards the algorithm ``converged'' and wiretaps other EVs' communication channels to obtain $\bm{\mathcal{C}}^{(\ell)}$ which will be used as an approximation of $\bm{\mathcal{C}}^*$. At any iteration $k$ after the $(\ell+1)$th iteration, the $i$th EV launches the stealthy attack by injecting
$\omega_1\nabla_{\bm{\mathcal{C}}_i}\mathcal{G}(\bm{\mathcal{C}}_i^{(k)})+2\omega_2\bm{I}_i^{\mathsf{T}}(\bm{\mathcal{C}}^{(k)}-\bm{\mathcal{C}}^{(\ell)})$ into its primal update. 

By implementing this stealthy attacking mechanism, the attacker can achieve personal benefit while manipulating the post-attack converged solution  to be close enough to the original optimal solution that the system operator is hard to detect any unusual anomaly. The attackers could tune the stealthy level $\omega_2$ in addition to their personal benefit attack power $\omega_1$ to balance the trade-off between gaining extra personal benefits and being more stealthy.

\noindent \textbf{Remark 1:} $\epsilon_{att}^*$ should not be too small, i.e., the approximation cannot be too accurate; otherwise, the algorithm would stop at full convergence before the $i$th EV launches the attack. 

\noindent \textbf{Remark 2:} The attacker must continuously wiretap other EVs' communication channels after the $\ell$th iteration.

\section{Simulation Results}
 The performance of different \emph{for-purpose} attacks will be demonstrated through simulations of controlling 500 EVs connected to a modified IEEE 13-bus test feeder \cite{DecentralizedEvSPDS_Liu}. Note that, Nodes 1 and 6 have no EV connected, and each of the other nodes is connected with 50 EVs equipped with level-2 chargers, i.e., $\tilde{P}_i=6.6$ kW. Battery capacities are uniformly distributed in $[18, 20]$ kWh. Initial and designated SOCs are uniformly distributed in $[0.3, 0.5]$ and $[0.7, 0.9]$, respectively. Primal and dual step sizes are empirically tuned to $\alpha_{i,k} = 2.8 \times 10^{-10}$ and $\beta_k = 1.8$, respectively.  The shrinking parameters are empirically chosen as $\tau_{\bm{\mathcal{C}}} = \tau_{\bm{\lambda}} = 0.974$. The maximum iteration number is set to $k_{max} = 25$. The convergence tolerance is chosen as $\epsilon = 1\times10^{-4}$. The voltage lower bound is set to $0.954$ p.u. The valley-filling period is from 19:00 to 8:00 the next day, which has been divided into 52 time periods with 15-minute lengths. The baseline
load data is scaled while collected from Southern California Edison \cite{baseload}. %All simulations are conducted in MATLAB + CVX on a Windows laptop with 3.3-GHz Intel Core i7 and 16-GB memory.
%In what follows, simulation results of the attack-free, non-stealthy  \emph{for-purpose} attacking, and stealthy \emph{for-purpose} attacking scenarios will be presented.
%We first analyze the EV charging problem without any attacks. Then, we implement the personal benefit attack without being stealthy. Finally, we present the \emph{stealthy for-purpose attack} and explore the impacts on network parameters like nodal voltages and total load.  $d_\lambda$ is selected as $d_\lambda = 5 \times 10^5$;
\subsection{Attack-free scenario}
\begin{figure}[!htb]
    \centering
    \includegraphics[width=0.4\textwidth, trim={0.1cm 0.4cm 0.0cm 0cm},clip]{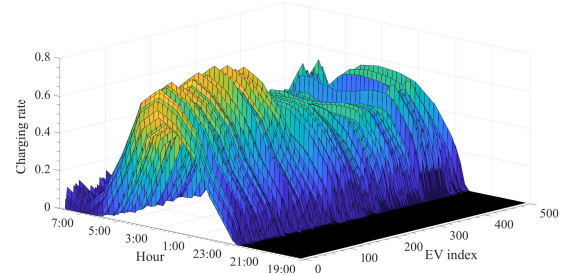}
    \caption{Charging profiles of all EVs in the attack-free scenario.}
    \label{no attack u}
\end{figure}

\begin{figure}[!htb]
    \centering  
\includegraphics[width=0.35\textwidth, trim={0cm 0.2cm 0.0cm 0.5cm},clip]{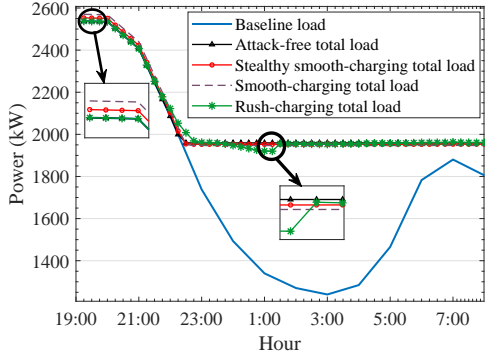}
    \caption{Baseline load and total load under different attacks.}
    \label{load profiles}
\end{figure}

\begin{figure}[!thb]\centering\includegraphics[width=0.485\textwidth, trim={0cm 0cm 0.0cm 0.2cm},clip]{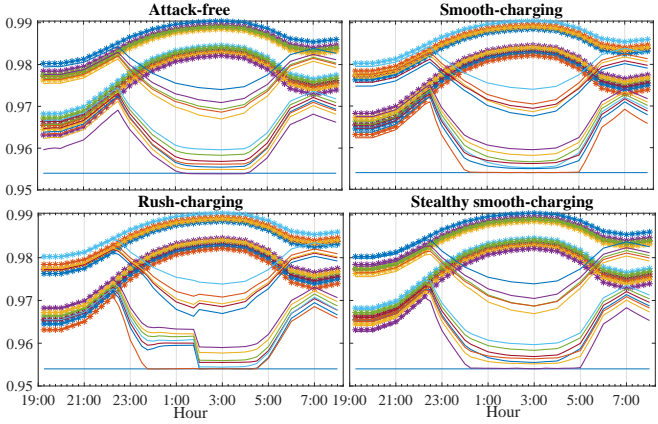}
    \caption{Nodal voltage magnitudes. Star-marked lines represent the baseline case, and solid lines represent the case with controlled EV charging loads.}
    \label{no attack v}
\end{figure}

By running SPDS to solve the attack-free problem in \eqref{main problem}, the charging profiles of all 500 EVs are shown in Fig. \ref{no attack u}.
It can be readily observed that all EVs stay idle before 22:30 and start charging after that to fill the valley. If EVs are grouped into four groups, i.e.,
Group 1 (1-200), Group 2 (201-300, 451-500), Group 3
(301-400), and Group 4 (401-450), according to their
geographic locations,
we can notice that the charging profiles of EVs in the same group
have the same trend. %Furthermore, it can be observed that EVs in Group 1 that are closer to the feeder's head than others have less impact on nodal voltage and can be pushed to have higher charging rates. On the other hand, the EVs in Group 4 that are further to the feeder's head will have a higher line impedance, and because of that, the network nodal voltages are much more sensitive to their charging powers.
The valley filing performance and the nodal voltage magnitudes under the attack-free scenario are shown in Fig. \ref{load profiles} and  Fig. \ref{no attack v}, respectively.   It can be observed that the total load profile becomes flat and stays at $1,958$ kW after 22:30, and all nodal voltage magnitudes are maintained above 0.954 p.u. 

\subsection{Smooth-charging attack} \label{smooth attack}
\begin{figure}[!htb]
   \centering
\includegraphics[width=0.4\textwidth, trim={0cm 0.4cm 0.0cm 0cm},clip]{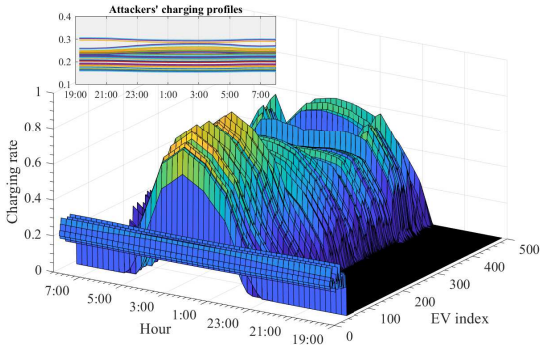}
    \caption{Charging profile of all EVs under smooth-charging attacks.}
    \label{smooth charging u}
    \vspace{-2mm}
\end{figure}

To better illustrate the smooth-charging attack, the first 50 EVs are selected as attackers. The self-interest attack power is set to $\omega_1=1\times10^5$ to make attacking impacts observable. Fig. \ref{smooth charging u}  shows the charging profiles of all EVs, where the attackers charge for all the time periods at almost a constant charging rate between $0.15$ to $0.3$. Meanwhile, the charging profiles of other EVs have no significant changes. Non-attacker EVs in the first Group  charge at higher rates from 00:00 to 4:00 in contrast to the attack-free scenario as they need to compensate for the 50 attackers in achieving  valley filling.

As shown in Fig. \ref{load profiles}, the valley filling objective can still be fulfilled under smooth-charging attacks. However, unlike the attack-free scenario, the controlled total load is slightly higher than the baseline load before 22:15 as the attackers start charging from the beginning. The flat value of the total load after 22:30 is $0.5\%$ lower than that of the attack-free scenario. The voltage behaviors under smooth-charging attacks are similar to those in the attack-free scenario, which can be found in Fig. \ref{no attack v}. If the system operator only monitors the nodal voltages, this attack is not making any suspicious impact.

%These results demonstrate that smooth-charging attacks can successfully smooth the attacker EVs' charging profiles while performing valley filling and maintaining nodal voltages. In practical cases, to stay stealthy, EVs launching the smooth-charging attack can tune down $\omega_1$ to have a smoother but less flattened charging profile.  

\subsection{Rush-charging attack}
\begin{figure}[!htb]
    \centering
\includegraphics[width=0.4\textwidth, trim={0cm 0.3cm 0.0cm 0cm},clip]{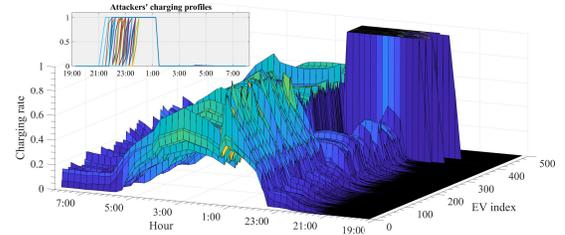}
    \caption{Charging profile of all EVs under rush-charging attacks.}
    \label{charging asap u}
\end{figure}

To better illustrate the rush-charging attack, 50 EVs in Group 4 are selected as attackers. The self-interest attack power is set to $\omega_1=1$, while $t_d=25$, $m=0.2$, and $M=1\times10^5$. Fig. \ref{charging asap u}  shows the charging profiles of all EVs in this scenario. It can be observed that the attackers in Group 4 start to charge at full power starting from 21:00 until they reach their $SOC_{des}$ before 2:00. The spike increase in the charging profiles of these 50 EVs led to lower EV charging rates in Groups 2 and 3 during the attacking period.

The valley-filling performance under rush-charging attack can be found in Fig. \ref{load profiles}, which presents the impaired performance. The total load is not entirely flattened, with the maximum value hitting $1,920$ kW while the final value being $1,961$ kW.  Unlike the smooth-charging attack, the voltage behaviors are very distinguishable in this case, as shown in  Fig. \ref{no attack v}. Though all the nodal voltages are still above 0.954 p.u., sudden drops and increases exist due to the rush charging.

% There is a drop in some nodal voltages while others have a slight increase which is related to the charging profiles represented in Fig. \ref{charging asap u}.  
%\begin{figure}[!htb]
%    \centering
    %\includegraphics[width=0.85\linewidth]{pscc2022_template_LaTeX/charge asap load crop.pdf}
    %\caption{Baseline load and total load under charging asap attack}
    %\label{charging asap load}
%\end{figure}

%\begin{figure}[!htb]
%    \centering
%    \includegraphics[width=0.85\linewidth]{pscc2022_template_LaTeX/charge asap voltage crop.pdf}
%    \caption{Nodal voltage magnitudes in the rush-charging scenario.}
%    \label{charging asap voltage}
%\end{figure} 

%Note that the attackers could have launched the rush-charging attack at the beginning of valley filling by choosing smaller $t_d$ and $m$, resulting in immediate full-power charging. However, by delaying the attacks, the attackers could conceal their attacks from the operator's view by eliminating a suspicious jump in total load. 

\subsection{Stealthy smooth-charging attack}
\begin{figure}[!htb]
    \centering
\includegraphics[width=0.4\textwidth, trim={0cm 0.2cm 0.0cm 0.05cm},clip]{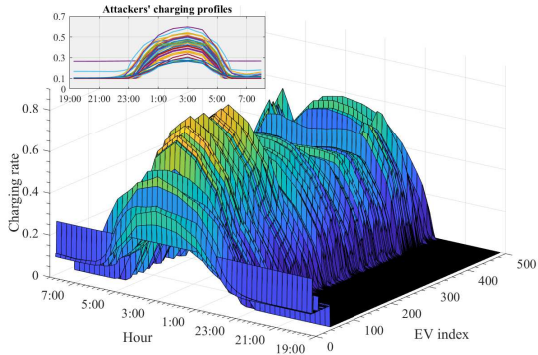}
    \caption{Charging profiles of all EVs under stealthy smooth-charging attacks.}
    \label{stealthy attack u}
\end{figure}

In this case, $\epsilon_{att}^*$ is set to $1.1\times10^{-4}$, leading the attacker to choose $\bm{\mathcal{C}}^{(19)}$ as the approximated optimal solution. We adopt the same setup for the smooth-charging attack of the first 50 EVs and set the stealthy attack power to $\omega_2=100$. The charging profiles of all EVs in this scenario can be found in Fig. \ref{stealthy attack u}, 
which shows smooth charging is realized for the first 50 EVs. The charging profiles of the attackers also follow the trend of other EVs in Group 1, which differentiates from the results shown in Fig. \ref{smooth charging u}.

The optimal values, i.e., the value of $\mathcal{F}(\cdot)$, under the attack-free, non-stealthy smooth-charging attack, and stealthy smooth-charging attack scenarios are $8.48866\times10^6$, $8.5089\times10^6$, and $8.5022\times10^6$, respectively, where the differences are unnoticeable. Let $\bm{\mathcal{C}}^*$ denote the attack-free optimal solution and define $\zeta=\|\hat{\bm{\mathcal{C}}}-\bm{\mathcal{C}}^* \|_2$ as an indicator for stealthiness where  $\hat{\bm{\mathcal{C}}}$ is the optimal solution under attacks. The values of $\zeta$ under the non-stealthy smooth-charging attack and stealthy smooth-charging attack scenarios are $13.02$ and $6.44$, respectively, indicating $49\%$ improvement in stealthiness.

%If $\bm{\mathcal{C}^\star}$ is the optimal solution of the attack-free scenario, Another analysis to show the improvement of the stealthiness is to evaluate $\|\bm{\mathcal{C}}-\bm{\mathcal{C}^\star}\|_2$; Which is equal to $13.02$ and $6.44$ in smooth-charging and stealthy smooth-charging attack, respectively. This proves that the deviation from the optimal solution has been reduced by $49\%$ in the stealthy case. 

%However, the smoothness of the charging profiles for the attackers has been reduced where for $i=1,2,\dots,50$ the $\|\bm{\mathcal{C}}_i\|_2$ changed from $11.19$ to $13.07$.

 The valley filling performance and nodal voltage behaviors are shown in Fig. \ref{load profiles} and Fig. \ref{no attack v}, respectively. Compared with the non-stealthy case, less total load deviation at the beginning can be observed. At 21:30, the total loads of the non-stealthy and stealthy scenarios are $2,570$ kW and $2,553$ kW, respectively, while the baseline load is $2,537$ kW, indicating $48\%$ lower load deviation.  Compared to that of the non-stealthy case, the nodal voltages in the stealthy case have around $57\%$ less deviation from the attack-free case. These results imply that the proposed stealthy attacking mechanism can effectively reduce the deviations from the truly optimal operation, thus improving attacking stealthiness. %With this very small difference, it is very hard for the control center to distinguish whether an attack is happening from a group of EVs. 

%\begin{figure}[!htb]
%    \centering
%    \includegraphics[width=0.85\linewidth]{pscc2022_template_LaTeX/v.jpg}
%    \caption{Nodal voltage magnitudes in the stealthy smooth-charging attack case.}
%    \label{stealthy attack voltage}
%\end{figure}

%It should be noted that the stealthy attack results can be different when the attacker choose different attack powers. For illustration, you can imagine changing the stealthy attack power $\omega_2$ from $0$ to $100$ will change the charging profile of the EVs from Fig.\ref{smooth charging u} to Fig.\ref{stealthy attack u}. 

\section{Conclusion}
This paper inaugurated \emph{for-purpose} algorithmic attacks that target general DMAO algorithms. By utilizing the \emph{for-purpose} algorithmic attacks,  algorithm participants can achieve self-interest purposes without affecting the algorithm convergence. Attack vectors with and without the stealthy feature were theoretically investigated and illustrated through a decentralized EV charging control problem. The efficacy of the proposed \emph{for-purpose} algorithmic attack was verified through EV charging control simulations. This paper is one of the first steps in bringing awareness of cyber attacks launched by DMAO participants and integrated into the algorithms. Comprehensive theoretical analyses will be provided in future work.

% we will exercise more developed self-interest algorithmic attacks and optimize the parameters associated with each attack to improve their gain and stealthiness further. 

%We implemented our proposed attack on a decentralized EV charging control problem associated with valley-filling, which is formulated based on the SPDS algorithm. It has been proved that the attackers can inject specific data into the updated iteration information and manipulate the algorithm to converge to the attackers' desired solution. Based on that, we propose two attack vectors, each representing a sample of a self-interest algorithmic attack. Further, we develop an optimization-based stealthy \emph{for-purpose} attacking mechanism to amplify the stealthiness of the deployed attacks. We verified the practicability of the proposed \emph{for-purpose algorithmic attacks} through simulations on a standard IEEE test system with realistic data. The results show the attackers can successfully deploy self-interest attacks while keeping low impacts on algorithm convergence, network nodal voltages, and total load. The final purpose of this study is to improve the security of the power grid and try to recognize possible manipulations in order to immune the grid furthermore. In future work, we will exercise more developed self-interest algorithmic attacks and optimize the parameters associated with each attack to improve their gain and stealthiness further.     
    
\bibliographystyle{IEEEtran}
\bibliography{Refrences}

\end{document}